\documentclass[aps,showkeys,showpacs,nofootinbib,floatfix]{revtex4}

\usepackage{amssymb}
\usepackage{amsfonts}
\usepackage{amsmath}
\usepackage{graphicx}
\usepackage{hyperref}
\usepackage{graphicx}
\begin{document}

\title{New asymptotic Anti-de Sitter solution with a timelike extra dimension in 5D relativity}
\author{Molin Liu}
\thanks{Corresponding author\\E-mail address: mlliu@xynu.edu.cn}
\author{Yingying Shi}
\author{Zonghua Zhao}
\author{Yu Han}
\affiliation{Institute for Gravitation and Astrophysics, College of Physics and Electronic Engineering, Xinyang Normal University, Xinyang, 464000, P. R. China}
\begin{abstract}
In 5D relativity, the usual 4D cosmological constant is determined by the extra dimension. If the extra dimension is spacelike, one can get a positive cosmological constant $\Lambda$ and a 4D de Sitter (dS) space. In this paper we present that, if the extra dimension is timelike oppositely, the negative $\Lambda$ will be emerged and the induced 4D space will be an asymptotic Anti-de Sitter (AdS). Under the minimum assumption, we solve the Kaluza-Klein equation $R_{AB} = 0$ in a canonical system and obtain the AdS solution in a general case. The result shows that an AdS space is induced naturally from a Kaluza-Klein manifold on a hypersurface (brane). The Lagrangian of test particle indicates the equation of motion can be geodesics if the 4D metric is independent of extra dimension. The causality is well respected because it is appropriately defined by a null higher dimensional interval. In this 5D relativity, the holographic principle can be used safely because the brane is asymptotic Euclidean AdS in the bulk. We also explore some possible holographic duality implications about the field/operator correspondence and the two-points correlation functions.
\end{abstract}



\maketitle
%\maketitle IS IGNORED %%%%%%%%%%%

\section{Introduction}\label{intro}
The idea that the world may have more than 4 dimensions is due to Kaluza, who realized that 5D manifold can unify Einstein's theory of general relativity (GR) with Maxwell's theory of electromagnetism \cite{Kaluza}. Then Einstein endorsed the idea, and a major impetus was provided by Klein, who connected it with quantum theory by considering a very small extra dimension \cite{Klein}. The 5D theory of Kaluza-Klein in a sense laid the foundation for the developments of modern theories such as 11D supergravity and 10D superstrings. At present 5D relativity is expanded to two modern versions: membrane theory and induced-matter theory (IMT) \cite{stm-reviews4}. Both theories build the physics in a 5D manifold by using a non-compactified extra dimension. In this paper we focus on the latter. In IMT the traditional gauge symmetry and particle content appearing in the standard model (SM) are embedded in a 5D flat space \cite{stm-reviews4,stm-reviews1,stm-reviews2,stm-reviews3}. There are many works studying on it, including the classical tests in solar system \cite{stm-apj1,stm-apj2}, the big bounce cosmology \cite{stm-bigbounce1,stm-bigbounce2}, the quasi-normal modes \cite{QNM-BH} and the thermodynamics \cite{entropy1-BH,entropy2-BH} of black hole, and so on. About the latest development of the 5D relativity, please refer to recent review \cite{stm-reviews4}. In a word, it cannot be ruled out through current observations.

Due to the fundamental rule of extra spatial dimension in string theory, the spacelike extra dimension is often adopted in higher dimensional gravity. In brane world, the brane is embedded in a Lorentzian multidimensional manifold by assuming a spacelike extra dimension \cite{ADD1,ADD2,RS1,RS2,RMaartens}. However, there is no a priori reason why extra dimensions must be spacelike. The timelike case can not be ruled out by the current observational constraints on the brane world \cite{GDvali}. This type of higher dimensional gravity can present us different features because of the timelike behaviour. For example, a natural cosmological bounce will appear and the FRW-type singularity is absent if the bulk's extra dimension is timelike \cite{YShtanov}. The hierarchy problem can also be reconciled with a correct cosmological expansion of the visible universe by extra timelike dimension in Randall-Sundrum brane \cite{MChaichian}. The effective 4D cosmological constant vanishes and avoids phenomenological difficulties from the matter instability by constructing one brane configuration in 6D space, in which one extra dimension is spacelike and the other is timelike \cite{ZBerezhiani}. The generic closed system of equations on a brane that describes its inner evolution is also obtained by considering the bulk spaces with spacelike/timelike extra dimension, and with/without $Z_2$ symmetry of reflection relative to the brane \cite{YVShtanov}. Hence, the spacelike and timelike extra dimensions are both accepted in higher dimensional space.

On the other hand, it is believed that the natural ground state is related to Anti-de Sitter (AdS) space in supergravity. AdS space thus gradually catches people's attention with the development of supergravity and string theories. In the classical thermodynamic system, the entropy is the characterization of the degree of freedom and its value is proportional to the volume of system (3D). However, when viewing from gravity, the entropy of black hole is proportional to the area of horizon (2D). This phenomenon indicates a 3D world to be the image of data stored on a 2D projection like a holographic image. The degree of freedom of a gravitational system could be described by that of a field in the boundary. t' Hooft therefore proposed that gravity has the characterization of holography \cite{tHooft} which is expanded subsequently by Susskind \cite{LSusskind}. Maldacena conjectured that the t' Hooft limit of 4D $\mathcal{N} = 4$ super-Yang-Mills at the conformal point is related to IIB strings \cite{JMMaldacena}. Under the background of $AdS_5 \times S^5$ the theory of IIB string is equivalent to that of (3+1)D super-Yang-Mills with $U(N)$ symmetric CFT living on the boundary. This conjecture is called AdS/CFT duality, which is regarded as an important achievement of holographic principle and is checked by many tests \cite{OAharony}. With the development of AdS/CFT duality \cite{Witten1,Gubser1,IRKlebano,DTSon}, people find it could be generalized to a more general situation: gauge/gravity duality, in which the theories of two sides need not to be supersymmetric. The current consensus over holographic duality is that a $(D + 1)$ gravitational theory possible is equivalent to a $D$ quantized field theory.

In the 5D relativity, the cosmological constant $\Lambda$ is determined by the extra dimension \cite{stm-reviews4,stm-reviews1,stm-reviews2,stm-reviews3}. The spacelike/timelike extra dimension will offer us a positive/negative $\Lambda$. In the literatures, much of the works focused on the spacelike case. However, people have recently found that the timelike extra dimension could be used to describe the de Broglie waves in the Einstein vacuum \cite{wavelikes1} or in the medium of radiation \cite{wavelikes2}. The waves in vacuum or radiation, which are compatible with 4D wave mechanics, are essentially 5D in nature because the classical 5D Klein-Gordon equations are respected well \cite{wavelikes3}. As far as we know, the AdS solution of 5D relativity has not been studied \cite{stm-reviews4}. If the timelike extra dimension can give us a negative cosmological constant, we can get the 3D Lorentzian space with one negative constant curvature. Motivated by these situations, we focus on the timelike extra dimension and try to build an AdS solution in a general case.

This paper is organized as follows. In section II, we present a new AdS solution to Kaluza-Klein equation by using a timelike extra dimension. In section III, we calculate the fundamental equations of motion by using the Lagrangian method. In section IV, we present a toy holographic duality implication. Section V is conclusion. Following the notation of Wesson \cite{stm-reviews2,stm-reviews3}, we absorb the constants $c$, $G$ and $h$ by a choice of units which renders their magnitude unity. Meanwhile we label 5D quantities with upper-case Latin letters ($A, B = 0, 1, 2, 3, 4$) and 4D quantities with lower case Greek letters ($\alpha, \beta = 0, 1, 2, 3$). If there is a chance of confusion between the 4D part of a 5D quantity and the 4D quantity as conventionally defined, we will use a hat to denote the former and the straight symbol to denote the latter.
\section{Anti-de Sitter solutions and negative cosmological constant}\label{AdSsolution}
In this section, we solve the 5D Einstein field equation $R_{AB} = 0$, and try to build a new class of AdS solutions. We first consider a smooth 5D space manifold which has an arbitrary metric tensor $g_{AB}$ satisfying the Kaluza-Klein equations $R_{AB} = 0$. The line element $d s$ of 5D is invariant under the group of general coordinate transformations. Then we adopt the canonical coordinates: $g_{\mu 4} = 0$ and $g_{44} = 1$ in this 5D manifold. In the literatures the canonical coordinate system is extensively adopted \cite{wavelikes1,cc-BMashhoon1,cc-PSWesson1,cc-HYLiu,cc-BMashhoon2}. The metric takes the form
$^{(5)} d s^2 = \hat{g}_{\mu\nu}(x; l) d x^{\mu} d x^{\nu} + d l^2$. It is possible to define a metric tensor $g_{\mu\nu} (x;l)$ satisfying $\hat{g}_{\mu\nu} = (l/L)^2 g_{\mu\nu}$. Under the canonical coordinates system, the 5D manifold  preserves the extra dimension with a distinct geometry structure and its 4D components then can be read directly from $^{(5)} d s^2$. Hence in a canonical coordinate system, we can initially write the 5D metric in the following form as
\begin{equation}\label{1-1metric}
d s^2 = \frac{l^2}{L^2} g_{\alpha\beta}(x^\alpha, l) d x^\alpha d x^\beta + d l^2,
\end{equation}
where the constant length $L$ preserves physical dimensions. In this canonical coordinate system, the extra dimensional coordinate lines are geodesics normal to the constant fifth dimensional hypersurfaces. Consequently, starting from a suitable initial hypersurface of $\hat{g}_{\mu\nu}(x; l)$, one can choose the extra dimensional coordinate lines to be the congruence of geodesic lines normal to the initial hypersurface. Along each such
line, the 5D interval corresponds to the canonical extra dimensional coordinate. The geometric construction is thus straightforward built under the canonical coordinate system, which could be described as the 5D analogue of the construction of the synchronous coordinate system in 4D space \cite{LDLandau}.

After choosing a canonical extra dimensional coordinate, the corresponding coordinates of 4D part could still be subjected to arbitrary transformations independent of extra dimension. One may recognize the 4D space part as the AdS metric with a negative cosmological constant $\Lambda = -3/L^2$ which means that the flat 5D space manifold corresponds to the empty AdS space. It turns out that the 4D space of the 5D metric in canonical coordinates is independent of extra dimension, the Kaluza-Klein equations reduce to the vacuum gravitational field equations with a negative cosmological constant. To prove this fact, one can consider the metric in canonical coordinates shown by Eq.(\ref{1-1metric}) and the components of the 5D Ricci tensor can be written as
\begin{eqnarray}
% \nonumber to remove numbering (before each equation)
^{(5)}R_{\mu\nu} &=& ^{(4)}R_{\mu\nu} - S_{\mu\nu},\\
^{(5)}R_{\mu 5} &=& A^{\alpha}_{\mu ;\alpha} -
\overset{\star}{\Gamma^{\alpha}_{\mu\alpha}}, \\
^{(5)}R_{55} &=& - \overset{\star}{A_{\alpha}^{\alpha}} - \frac{2}{l} A_{\alpha}^{\alpha} - A_{\alpha\beta}A^{\alpha\beta},
\end{eqnarray}
where the semicolon denotes the usual covariant differentiation in 4D and the star denotes the partial derivative with respect to extra dimension $l$, e.g. $\overset{\star}{A_{\alpha}^{\alpha}} = \partial A_{\alpha}^{\alpha}/\partial l$. $S_{\mu\nu}$ is a symmetric tensor shown by
\begin{equation}\label{Smunu}
S_{\mu\nu} = \frac{l^2}{L^2} \left[2A_{\mu}^{\alpha} A_{\nu\alpha} - \overset{\star}{A_{\mu\nu}} - \left(\frac{4}{l} + A_{\alpha}^{\alpha}\right)A_{\mu\nu}\right] - \frac{1}{L^2} \left(3 + l A_{\alpha}^{\alpha}\right) g_{\mu\nu},
\end{equation}
where $A_{\alpha\beta} = \frac{1}{2} \overset{\star}{g_{\alpha\beta}}$ with $\overset{\star}{g_{\alpha\beta}} = \partial g_{\alpha\beta}/ \partial l$. $^{(4)}R_{\mu\nu}$ and $\Gamma^{\mu}_{\nu \rho}$ are the 4D Ricci tensor and the connection coefficients by $g_{\alpha\beta}$. The 4D part of Eq.(\ref{1-1metric}) satisfies following Einstein equation,
\begin{equation}\label{4Deinsteineq}
^{(4)} R_{\mu\nu} - \frac{1}{2}\ ^{(4)}R g_{\mu\nu} + \Lambda g_{\mu\nu} = \frac{8\pi G}{c^4} T_{\mu\nu}.
\end{equation}
Therefore, the energy-momentum tensor is written as
\begin{equation}\label{emten}
T_{\mu\nu} = \frac{c^4}{8 \pi G} \left[S_{\mu\nu} - \frac{1}{2} g_{\mu\nu} (S - 2\Lambda)\right].
\end{equation}
If the 4D metric $g_{\alpha\beta}$ is independent of extra dimension, we can have $A_{\alpha\beta} = 0$ which could ensure the off diagonal $^{(5)}R_{\mu 5}$ and the fifth diagonal $^{(5)}R_{5 5}$ vanish. The 4D Ricci tensors satisfy $^{(4)}R_{\mu\nu} = S_{\mu\nu} = - 3 g_{\mu\nu}/L^2$. Then according to the vacuum condition $T_{\mu\nu} = 0$, we can obtain a negative cosmological constant $\Lambda = - \frac{3}{L^2}$.

On the other hand, if we check the 4D Einstein tensor $G_{\alpha\beta}$ expressed by
\begin{eqnarray}
\nonumber G_{\alpha\beta} &=& \frac{1}{2 L^2} \left[l^2 \overset{\star\star}{g}_{\alpha\beta} - l^2 g^{\lambda\mu} \overset{\star}{g}_{\alpha\lambda} \overset{\star}{g}_{\beta\mu} + 4 l \overset{\star}{g}_{\alpha\beta} + \frac{l^2}{2} g^{\mu\nu}\overset{\star}{g}_{\mu\nu} \overset{\star}{g}_{\alpha\beta}\right]\\ &-& \frac{1}{2 L^2} \left[6 + 2 l g^{\mu\nu} \overset{\star}{g}_{\mu\nu} + \frac{l^2}{4}\overset{\star}{g}^{\mu\nu}\overset{\star}{g}_{\mu\nu} + \frac{l^2}{4} \left(g^{\mu\nu}\overset{\star}{g}_{\mu\nu}\right)^2\right] g_{\alpha\beta},\label{1-galhabelta}
\end{eqnarray}
we can find this equation could be just $G_{\alpha\beta} = - 3 g_{\alpha\beta} / L^2$ if the extra dimension is independent of $l$. It is also indicates that the 4D Einstein's equations reduced from 5D metric (\ref{1-1metric}) has a negative constant $\Lambda = -3/L^2$. Hence,
an $AdS_4$ is embedded naturally in 5D Ricci flat space. Viewing from the point of modern Kaluza-Klein theory, the negative cosmological constant of gravity is a kind of artifact produced by reducting from 5D to 4D. Therefore, the AdS Schwarzschild black hole of 5D relativity are given by,
\begin{equation}\label{15dadssolution}
d s^2 = \frac{l^2}{L^2}\left[f(r) d t^2 - \frac{d r^2}{f(r)} - r^2 (d x^2 + d y^2)\right] + d l^2,
\end{equation}
where the metric function is $f(r) = \frac{r^2}{L^2} - \frac{M}{r}$, the 4D cosmological constant is $\Lambda = -3/L^2$.
It should be noticed that this kind of timelike extra dimension is widely used in 5D relativity and does not have the physical nature of a time \cite{stm-reviews4,wavelikes1,wavelikes2,wavelikes3}. After checked by computer the solution (\ref{15dadssolution}) is indeed the exact solution to Kaluza-Klein field equation $R_{AB} = 0$.

On the brane identified by the hypersurface of $l = constant$ in the 5D bulk, one can immediately get a standard planar 4D AdS Schwarzschild black hole shown by
\begin{equation}\label{add-4D-AdS-Schwarz}
d s^2 = f(r) d t^2 - \frac{d r^2}{f(r)} - r^2 (d x^2 + d y^2),
\end{equation}
where the constant coefficient $L^2/l^2$ is absorbed into the 4D line element $d s^2$. It agrees with the general case which has been extensively studied in holographic principle to build an AdS/CFT superconductor \cite{Horowitz_PRD,Hartnoll_PRL}, in which the opposite notations $(-, +, +, +)$ are adopted in the metric. The horizon in Eq.(\ref{15dadssolution}) is $r_H = M^{1/3} L^{2/3}$ obtained by the null condition $n_\mu n^\mu = 0$. The Hawking temperature therefore is given by
\begin{equation}\label{0-temprature}
T = \frac{1}{4\pi}\frac{\partial f(r)}{\partial r} \bigg{|}_{r_H} = \frac{M L^2 + 2 r_H^3}{4 \pi L^2 r_H^2} = \frac{3M^{1/3}}{4 \pi L^{4/3}}.
\end{equation}
The heat capacity defined by $C = \frac{\partial M}{\partial T}$ takes the positive value $1/4 \pi r_H^2$ which ensures the thermodynamic stability. Hence a Schwarzschild black hole could be in thermal equilibrium in the AdS space and the AdS plays the role of a box. It also should be noted that the positive cosmological constant can be obtained by a spacelike extra dimension, and the black hole solutions with cosmological constant are extensively studied in literatures \cite{stm-bigbounce1,stm-bigbounce2,cc-BMashhoon1,cc-HYLiu,cc-BMashhoon2} which are completely different from above AdS solution (\ref{15dadssolution}) with a timelike extra dimension.

It is known that the usual 4D AdS space can be embedded in a higher 5D flat space with a timelike extra dimension. This flat space is non-Euclidean and non-Minkowski because the sign of metric is $(+, +, -, -, -)$ and the sign difference is $-1$. These points are very like these of the solutions (\ref{1-1metric}) or (\ref{15dadssolution}). Here, we can make this analogy and show the 4D metric of AdS from the 5D Ricci flat space (\ref{15dadssolution}). About the detail explanations of usual 4D AdS space embedded in a 5D flat one, please see Ref.\cite{Hawking,liangzhou}.

By adopting the coordinate transformations: $ T = \frac{l}{L} f^{1/2}(r) t, Z = \frac{l}{L} \int \frac{d r}{f (r)}, X = \frac{l r}{L} x, Y = \frac{l r}{L} y,  W =  l$, the 5D Ricci flat space (\ref{15dadssolution}) is rewritten as
\begin{equation}\label{add-flat-metric}
d s^2 = d T^2 - d Z^2 - d X^2 - d Y^2 + d W^2,
\end{equation}
which is just the 5D flat manifold ($\mathbb{R}^5$, $\tilde{\eta}_{ab}$) used to present the usual 4D AdS space in many textbooks such as \cite{Hawking,liangzhou}. Following the traditional method, we can define a 4D hypersurface $M^{4*}$ by the following equation
\begin{equation}\label{add-flat-metric2}
T^2 + W^2 - Z^2 - X^2 - Y^2 = l_0^2,
\end{equation}
where $l_0$ is a positive real constant. Then if we use $g_{ab}^{*}$ to denote the induced metric on $M^{4*}$ from $\tilde{\eta}_{ab}$, we can get the Riemann tensor $R_{abc}^{e}$ satisfied $R_{abcd} = 2 K g^*_{c[a}g^*_{b]d}$ where $R_{abcd} = g_{de} R_{abc}^{e}$ and $K = -l_0^{-2}$. Hence, $g_{ab}^{*}$ is a constant curvature metric and is the solution of vacuum Einstein equation with negative cosmological constant $\Lambda = -3 l_0^{-2}$. Specifying $X$, $Y$ and $Z$ by $X_1$, $Y_1$ and $Z_1$ in Eq.(\ref{add-flat-metric2}) respectively, we can have a 2D round equation about $(T, W)$ shown by
\begin{equation}\label{add-flat-metric3}
T^2 + W^2 = l_0^2 + X_1^2 + Y_1^2+ Z_1^2 > 0.
\end{equation}
Clearly, the intersection of $M^{4*}$ and the above planar $(T, W)$ is $S^1$, and therefore the topology of AdS space is $\mathbb{R}^3 \times S^1$.

Furthermore, in order to more clearly present the 4D AdS metric, we adopt the spherical symmetry coordinate system ($t'$, $r'$, $\theta$, $\varphi$) on the 4D hypersurface $M^{4*}$,
\begin{eqnarray}
% \nonumber to remove numbering (before each equation)
\label{add-T}  T &=& (l_0^2 + r'^2)^{1/2} \sin (l_0^{-1} t'), \\
\label{add-W}  W &=& (l_0^2 + r'^2)^{1/2} \cos (l_0^{-1} t'), \\
\label{add-X}  X &=& r' \sin \theta \cos \varphi, \\
\label{add-Y}  Y &=& r' \sin \theta \sin \varphi, \\
\label{add-Z}  Z &=& r' \cos \theta,
\end{eqnarray}
where $0< t' < 2\pi l_0$, $0< r' <\infty$, $0<\theta<\pi$, $0<\varphi<2\pi$. It is easy to verify that Eqs.(\ref{add-T})-(\ref{add-Z}) satisfy the hypersurface equation (\ref{add-flat-metric2}). Substituting the differential forms of Eqs.(\ref{add-T})-(\ref{add-Z}) into the metric (\ref{add-flat-metric}), we can obtain the known 4D AdS space as,
\begin{equation}\label{add-inducedlineelem}
(d s^*)^2 = -\left(1+\frac{r'^2}{l_0^2}\right) d t'^2 + \left(1+\frac{r'^2}{l_0^2}\right)^{-1} d r'^2 + r'^2(d \theta^2 + \sin^2\theta d \varphi^2),
\end{equation}
where the curvature of $M^{4*}$ is negative constant $K = -l_0^{-2}$. Except for the coordinate singularity, the full AdS space ($M^{4*}, g^{*}_{ab}$) will be covered by the coordinates ($t'$, $r'$, $\theta$, $\varphi$).
\section{The equations of motion}\label{EoM}
Next we need to check whether the motion of particle is geodesics in this new solution? In order to solve this problem, we calculate the Lagrangian of test particle and try to get the fundamental geodesic equation. The coordinate in Eq.(\ref{1-1metric}) is constructed by taking a 4D hypersurface in the 5D manifold in which the lines normal to this hypersurface serves as the extra coordinate. These lines are geodesics and proper length along them is the extra coordinate $x^4 = l$. This method of constructing a coordinate system is the 5D analog of how the synchronous coordinate system of general relativity is set up in 4D \cite{LDLandau}. It is certainly possible that this coordinate system breaks down if the coordinate lines in the fifth dimension cross. To this regard, apart from the pathological situations, the coordinates are always admissible within a finite interval along the fifth dimension. The equations of motion could be given by minimizing the distance between two points in 5D through the variation $\delta [\int d s] = 0$. The path in 5D is described by $x^{\alpha} = x^{\alpha}(l)$ and $l = l(\lambda)$ with an affine parameter $\lambda$ along the path. This relation can be rewritten as $\delta [\int \mathcal{L}\ d \lambda] = 0$ and the Lagrangian of particle shown by
\begin{equation}\label{Lagrangian}
\mathcal{L} \equiv \frac{d s}{d \lambda} = \left[\frac{l^2}{L^2}g_{\alpha\beta} \frac{d x^\alpha}{d \lambda}\frac{d x^\beta}{d \lambda} + \left(\frac{d l}{d \lambda}\right)^2\right]^{1/2}.
\end{equation}
Like the usual GR, the momenta are given by
\begin{eqnarray}\label{momenta}
\label{palpha} P_{\alpha} &=& \frac{\partial \mathcal{L}}{\partial(d x^{\alpha}/d \lambda)} = \left(\frac{l^2}{L^2}\right) \frac{g_{\alpha\beta}u^{\beta}}{\xi^{1/2}}, \\
\label{pl} P_{l} &=& \frac{\partial \mathcal{L}}{\partial(d l/d \lambda)} = \frac{u^{l}}{\xi^{1/2}},
\end{eqnarray}
where $u^{\alpha} = d x^{\alpha}/d \lambda$ and $u^l = d l/d \lambda$ are the velocities, and $\xi \equiv \mathcal{L}^2 = (l^2/L^2)g_{\alpha\beta}u^{\alpha}u^{\beta} + (u^l)^2$. Then according to Euler equation, the equations of motion are given by
\begin{equation}
% \nonumber to remove numbering (before each equation)
\frac{d P_{\alpha}}{d \lambda} = \frac{\partial \mathcal{L}}{\partial x^{\alpha}}, \ \ \ \ \frac{d P_{l}}{d \lambda} = \overset{\star}{\mathcal{L}}.
\end{equation}
By using above momenta (\ref{palpha}) and (\ref{pl}), we can get the corresponding equations of motion,
\begin{eqnarray}
% \nonumber to remove numbering (before each equation)
\label{eqmotion4-1} && \frac{d}{d\lambda}\left(g_{\alpha\beta}u^{\beta}\right) + \frac{2}{l}\left(\frac{d l}{d \lambda}\right)g_{\alpha\beta}u^{\beta} - g_{\alpha\beta}u^{\beta}\left(\frac{1}{2 \xi}\right)\frac{d \xi}{d \lambda} - \frac{1}{2}\frac{\partial g_{\beta\gamma}}{\partial x^{\alpha}} u^{\beta} u^{\gamma} = 0,\\
\label{eqmotion5-1} && \frac{d^2 l}{d \lambda^2} - \left(\frac{1}{2\xi}\right)\frac{d \xi}{d \lambda}\frac{d l}{d \lambda} - \frac{l}{L^2}g_{\alpha\beta}u^{\alpha}u^{\beta} - \frac{1}{2}\left(\frac{l^2}{L^2}\right)
\overset{\star}{g}_{\beta\gamma}u^{\beta} u^{\gamma} = 0.
\end{eqnarray}
In order to keep the usual 4D conservation law about the velocity, namely $g_{\alpha\beta} \frac{d x^{\alpha}}{d \lambda} \frac{d x^{\beta}}{d \lambda} = 1$, we can choose the affine parameter as $\lambda = s$ where $d s^2 = g_{\alpha\beta} d x^{\alpha} d x^{\beta}$ is the just 4D part of 5D metric (\ref{15dadssolution}). Then the equations of motion (\ref{eqmotion4-1}) and (\ref{eqmotion5-1})are rewritten as new forms as follows,
\begin{eqnarray}
% \nonumber to remove numbering (before each equation)
\label{eqmotion4-2} \frac{d u^{\mu}}{d s} + \Gamma^{\mu}_{\beta\gamma}u^{\beta}u^{\gamma}&=& -g^{\mu\alpha} \overset{\star}{g}_{\alpha\beta} \frac{d l}{d s} u^{\beta} + \left(\frac{1}{2\xi} \frac{d \xi}{d s} - \frac{2}{l}\frac{d l}{d s}\right) u^{\mu}, \\
\label{eqmotion5-2} \frac{d^2 l}{d s^2} - \left(\frac{1}{2\xi}\right)\frac{d \xi}{d s} \frac{d l}{d s}&=& \frac{l}{L^2} + \frac{1}{2}\left(\frac{l^2}{L^2}\right) \overset{\star}{g}_{\beta\gamma} u^{\beta}u^{\gamma},
\end{eqnarray}
where the 4D relationship $g_{\alpha\beta} u^{\alpha} u^{\beta} = 1$ is used. It needs to be noticed that we consider the case that 4D $g_{\alpha\beta}$ is generally dependent on the normal 4D coordinates ($x^{\mu}$) and the extra dimension $l$ and we can have following differential relation as
\begin{equation}\label{2-4d-5d}
\frac{d g_{\alpha\beta}}{d s} = \frac{\partial g_{\alpha\beta}}{\partial x^{\mu}} \frac{d x^{\mu}}{d s} + \overset{\star}{g}_{\alpha\beta}\frac{d l}{d s}.
\end{equation}
Here, $\Gamma_{\beta\gamma}^{\mu}$ is the usual 4D Christoffel symbol of the second kind shown by
\begin{equation}\label{Christoffel symbol}
\Gamma_{\beta\gamma}^{\mu} = \frac{1}{2} g^{\mu\alpha} \left(g_{\alpha\beta,\gamma} + g_{\alpha\gamma,\beta} - g_{\beta\gamma,\alpha}\right).
\end{equation}
Then after eliminating parameter $\xi$ by $\frac{d \xi}{d s} = 2\left(\frac{l}{L^2 } + \frac{d l}{d s} \frac{d^2 l}{d s^2}\right)$ through former definition of $\xi$, Eq.(\ref{eqmotion5-2}) could be rewritten as
\begin{equation}\label{2-5d-motion2}
\frac{d^2 l}{d s^2} - \frac{1}{l} \left(\frac{d l}{d s}\right)^2 = \left[\frac{l^2}{L^2} + \left(\frac{d l}{d s}\right)^2\right] \left[\frac{1}{l} + \frac{1}{2}
\overset{\star}{g}_{\beta\gamma} u^{\beta} u^{\gamma}\right],
\end{equation}
which can help us to simplify Eq.(\ref{eqmotion4-2}) and obtain the final 4D equations of motion
\begin{equation}\label{geodesiceq}
\frac{d^2 x^\mu}{d s^2} + \Gamma^{\mu}_{\alpha\beta} \frac{d x^\alpha}{d s}\frac{d x^\beta}{d s} = F^{\mu}.
\end{equation}
Here, $F^{\mu}$ is the extra force experienced by a test particle per unit proper mass and is expressed by \cite{cc-PSWesson1},
\begin{equation}\label{force}
F^{\mu} = \left(-g^{\mu\alpha} + \frac{1}{2} \frac{d x^{\mu}}{d s}\frac{d x^{\alpha}}{d s}\right) \frac{d l}{d s}\frac{d x^{\beta}}{ds} \overset{\star}{g}_{\alpha\beta}.
\end{equation}
This kind of extra force with different forms will appear in many solutions in canonical system (see Refs.\cite{stm-reviews2,stm-reviews3}). Comparing with the spacelike case, we can interestingly find the equations of motion in 4D have the same forms no matter whether the extra dimension is spacelike or timelike. However, when comparing with spacelike case \cite{cc-BMashhoon2}, the geodesic equation of timelike extra dimension has a completely different form as
\begin{equation}\label{fifthgeo}
\frac{d^2 l}{d s^2} -\frac{2}{l} \left(\frac{d l}{d s}\right)^2 - \frac{l}{L^2} = \frac{1}{2} \left[\frac{l^2}{L^2} + \left(\frac{d l}{d s}\right)^2\right] \frac{d x^{\alpha}}{d s} \frac{d x^{\beta}}{d s} \overset{\star}{g}_{\alpha\beta}.
\end{equation}
Moreover, Eq.(\ref{fifthgeo}) can be solved completely and the extra dimension is chosen as
\begin{equation}\label{fifthgeo-so}
l = l_0 \sec \left(\frac{s - s_0}{L}\right),
\end{equation}
where we are restricting to the 5D null paths, and $s/c$ is the proper time along the geodesic path of the particle. Interestingly, this result is completely different from the case of spacelike extra dimension \cite{cc-BMashhoon1} in which the extra dimension has the form of Hyperbolic functions, i.e. $l = l_0\text{sech} (\frac{s-s_0}{L})$. Obviously, the timelike extra dimension could give us a different content when comparing with the case of spacelike extra dimension. In fact, this result can also be justified by a transformation $l \rightarrow i l$, thereby Hyperbolic functions can be derived from trigonometric identity, namely $\text{sech} z = \sec iz$ \cite{Abramowitz}.

Meanwhile, the nonconservation relationship $F_{\mu}u^{\mu} \neq 0$ arises in 5D metric with canonical coordinates. It is distinctive because in usual 4D GR the timelike motions with known forces must obey equations of motion in the form of $D u^{\mu}/d s = F^{\mu}$ with $F_{\mu}u^{\mu} = 0$ which could be obtained by the condition of normalization involving the well-informed 4D velocity $u^{\mu}u_{\mu} = 1$. Conversely, the extra force (\ref{force}) possible is non 4D in origin. To check this, we can decompose $F^{\mu}$ into the sum of a component $N^{\mu}$ normal to the 4-velocity of the particle and another component $P^{\mu}$ parallel to it, which are shown by,
\begin{eqnarray}
% \nonumber to remove numbering (before each equation)
F^{\mu} &=& N^{\mu} + P^{\mu},\\
N^{\mu} &=& (-g^{\mu\alpha} + u^{\mu}u^{\alpha})u^{\beta}\overset{\star}{g}_{\alpha\beta}
\frac{d l}{d s},\\
P^{\mu} &=& - \frac{1}{2} u^{\mu} u^{\alpha}u^{\beta} \frac{d l}{d s} \overset{\star}{g}_{\alpha\beta}.
\end{eqnarray}
It becomes very clear that the normal $N^{\mu}$ could be due to ordinary 4D forces since it obeys $N^{\mu} u_{\mu} = 0$ by construction. However, the parallel component $P^{\mu}$ has no 4D analog because $P^{\mu} u_{\mu} \neq 0$ and $P^{\mu} = u^{\mu} F^{\alpha} u_{\alpha}$ in general case. The anomalous extra force is therefore a consequence of the existence of the extra dimension. Hence, the acceleration is not orthogonal to the velocity of the particle because there is relationship $g_{\mu\nu} F^{\mu} d x^{\nu}/d s \neq 0$ and $g_{\mu\nu}$ depends on the extra dimension $l(s)$. However, if the metric is independent of $l$, we can have $\overset{\star}{g}_{\alpha\beta} = 0$. The path of test particle in 4D is a geodesic shown by formula (\ref{geodesiceq}) with $F^{\mu} = 0$. Like the cases of spacelike extra dimension contained the canonical 1-body solution \cite{cc-BMashhoon2} and the canonical inflationary solution \cite{stm-bigbounce1}, $P^{\mu}$ is along the positive direction defined by $u^{\mu}$, which is also determined by the property of timelike extra dimension. This result is independent of whether the extra dimension is spacelike or timelike.

Then we need to discuss the problem of causality. The extra dimension is used commonly to be spacelike, whereas it is also can be timelike in many literatures in higher dimensional theories \cite{stm-reviews1,stm-reviews2,stm-reviews3}. The timelike or spacelike extra dimension does not cause any problem with causality \cite{stm-reviews1}. The 5D line element (\ref{1-1metric}) can be rewritten as $^{(5)} d s^2 = \frac{l^2}{L^2} d s^2 + d l^2$ where $d s^2$ is the 4D line element. Hence, the causality is logically defined by the 5D null paths given by $^{(5)} d s^2 = 0$ \cite{wavelikes1,wavelikes2,wavelikes3,stm-reviews3}. The conventional 4D paths for the photon and massive particle can be given by 4D interval or proper time by $d s^2 \geq 0$. When the extra dimension $l$ is timelike, $\Lambda < 0$ and the null geodesics are oscillatory. Therefore, $^{(5)} d s^2 = 0$ defines the 5D causality. This aspect of the situation is compatible with the condition $^{(4)} d s^2 = 0$ for causality as defined in 5D.

In the last, The 4D field equations derived here reduce to $G_{\alpha\beta} = \Lambda g_{\alpha\beta}$ only if the metric is independent of the extra dimension $l$. Thus the induced cosmological constant $\Lambda$ comes from the extra dimension $l$ in this 5D Ricci flat space. In the last, we need to show the particle geodesics reduce to known solutions when
the higher dimensional space is Minkowski. If the 5D space is Minkowski, the sign of metric is $(+, -, -, -, -)$ and the sign difference is $-3$. The extra dimension could be changed from timelike one to spacelike one by a transformation $l \longrightarrow i l$. Hence the 5D space (\ref{1-1metric}) with spacelike extra dimension will reduce to Minkowski when the metric $g_{\alpha\beta}$ is independent of $l$ and reduces to a usual 4D Minkowski metric $\eta_{\alpha\beta}$. In this way, the Lagrangian (\ref{Lagrangian}) of particle will reduce to a simple form as
\begin{equation}\label{add-2-lagrangian}
\mathcal{L} \equiv \frac{d s}{d \lambda} = \left[\eta_{\alpha\beta} \frac{d x^\alpha}{d \lambda}\frac{d x^\beta}{d \lambda} - \frac{d l}{d \lambda}\frac{d l}{d \lambda}\right]^{1/2} = \left[\eta_{AB} \frac{d x^A}{d \lambda}\frac{d x^B}{d \lambda}\right]^{1/2},
\end{equation}
where $\eta_{AB}$ is the 5D Minkowski metric, i.e. $\eta_{AB} = \text{diagonal}\ \ (+1,\ -1,\ -1,\ -1,\ -1)$.
Then in this way, the geodesics equations (\ref{eqmotion4-2}) and (\ref{eqmotion5-2}) can well be merged into the usual 5D Minkowski case as,
\begin{equation}\label{add-4d-geodesiceq}
\frac{d^2 x^A}{d s^2} + \Gamma^{A}_{B C} \frac{d x^B}{d s}\frac{d x^C}{d s} = 0,
\end{equation}
where the $\Gamma^{A}_{BC}$ is defined by usual 5D Minkowski metric $\eta_{AB}$. So we can say that if the extra dimension is spacelike, the particle geodesics reduce to the known solutions when the higher dimensional space is Minkowski.
\section{A toy model of holographic duality in 5D relativity}\label{holodual}
\begin{figure}[!htb]
\centering
% Requires \usepackage{graphicx}
\includegraphics[width=3.0 in]{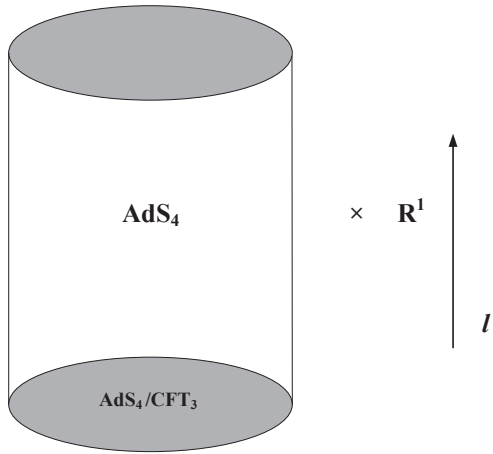}\\
\caption{The $AdS_4/CFT_3$ on the 4D hypersurface $M^{4*}$ in the 5D Ricci flat space with a non-compactification extra dimension $l$. Here, the hypersurface equation is $T^2 + W^2 - Z^2 - X^2 - Y^2 = -3/\Lambda$. The induced metric $g_{\alpha\beta}^{*}$ of the hypersurface $M^{4*}$ is defined by Eq.(\ref{add-inducedlineelem}) in the coordinate system ($t$, $r$, $\theta$, $\varphi$).}\label{Fig1}
\end{figure}

In this section, we will check its Euclidean version and study its boundary field defined by the gravity in the bulk (see Fig.\ref{Fig1}). Although there are various boundary observations that need to be considered, we focus on the simple two-points correlation functions as being a toy implication. The operators dual to scalars and two-points correlation functions of scalar operators are performed in the standard route.

Considering the brane of the 5D relativity solution (\ref{15dadssolution}) is asymptotic AdS. The extra dimension does not have the physical nature of a time \cite{wavelikes1}, we thus may possibly use the holographic principle on these surfaces. According to the 5D space (\ref{15dadssolution}), we can find the parts of ($t$, $r$, $x$, $y$) dependents the extra dimension, i.e. $l^2/L^2$. Therefore the holographic principle need to consider the full 5D picture. But because the 5D space is Ricci flat and the AdS is induced on the 4D hypersurface $M^{4*}$, the right correspondence should be between 4D AdS and 3D CFT. In another words, in this model an $AdS_4/CFT_3$ exists on the 4D hypersurface $M^{4*}$ which is embedded in the 5D Ricci flat space with a non-compactification or compactification extra dimension $l$. Based on the analysis about the induced 4D AdS metric $g_{\alpha\beta}^{*}$ in section \ref{AdSsolution}, we can know that the $AdS_4/CFT_3$ adopted here is different from the usual $AdS_5/CFT_4$. The correspondence of $AdS_4/CFT_3$ is illustrated in Fig.\ref{Fig1}.
\subsection{field/operator corresponding}\label{sub-fieldoperator}
We first calculate the scalar field near the Breitenlohner-Freedman boundary. After adopting a coordinate transformation $u = 1/r$, the Euclidean version of metric (\ref{15dadssolution}) reduces to
\begin{equation}\label{1-boundarymetric}
d s^2 = \frac{l^2}{L^2 u^2} \left[\frac{d t^2}{L^2} + L^2 d u^2 + \left(d x^2 + d y^2\right)\right] + d l^2,
\end{equation}
where the limit $f(u)\rightarrow 1/L^2u^2$ is used.

Before the formal calculation, it is necessary to explain a situation of that the AdS part of 5D metric (\ref{1-boundarymetric}) contains the extra dimensional term, i.e. a coupling factor $l^2/L^2$ before the parts of $(t, u, x, y)$. Therefore, we can not only simple consider the 4D part insider the brackets. Meanwhile, through the Fig.\ref{Fig1} we also can find the extra dimension lives the whole space including $AdS_4$ and $CFT_3$. So as a conservative consideration, we proceed from the original 5D spaces (\ref{1-boundarymetric}) to calculate 4D quantum field at the boundary of $AdS_4$. The 3D CFT used to implement $AdS_4/CFT_3$ is actually included in the 4D quantum field at the boundary of $AdS_4$. The real duality is the $AdS_4/CFT_3$, not the $AdS_5/CFT_4$ \footnote{Please note that there is no $AdS_5$ in metric (\ref{1-1metric}) because the 5D space is Ricci flat.}.

Under above gravitational background a scalar field $\psi$ is considered to exist in this bulk and $\psi_0$ is adopted as the restriction of $\psi$ to the boundary of AdS. We assume the bulk field $\psi$ is coupled to an operator $\mathcal{O}_\psi$ by $\int_{S^4}\psi_0\mathcal{O}_\psi$ on the boundary $\psi_0$. Then we consider the scalar field on this background, the action of the scalar can thus be given by
\begin{equation}\label{1-scalarfield}
S_{\psi} = \frac{1}{2} \int d^5 x \sqrt{g} \left(g^{AB}\partial_A \partial_B \psi + m^2 \psi^2\right),
\end{equation}
where $m$ is the scalar mass. In order to analyse the boundary field, we use a Fourier expansion $\psi = \phi (u, k^\mu) e^{ik_\mu x^\mu}$ with $k_\mu = (-\omega, \vec{k})$. If we consider the rotational symmetry along the spatial direction, we can get $k_1 = k$ and $k_i = 0$ when $i\neq 1$. Hence, the linearized field equation for $\phi$ is shown by
\begin{equation}\label{1-motioneq}
\frac{1}{\sqrt{g}} \partial_u \left(\sqrt{g} g^{uu} \partial_u\right) \phi - g^{tt} \omega^2 \phi - g^{xx} k^2 \phi - m^2 \phi = 0,
\end{equation}
which can be solved with fixed boundary condition at boundary $\psi_0$.
Submitting the metric (\ref{1-boundarymetric}) into above motion equation, one can get a solvable form of equation about $\phi$ as
\begin{equation}\label{1-boundary-phi}
u^2 \frac{\partial^2 \phi}{\partial u^2} - 2 u \frac{\partial \phi}{\partial u} - \left[m^2 l^2 + (L^2 k^2 + L^4 \omega^2) u^2\right] \phi = 0.
\end{equation}
Obviously, the extra dimension $l$ effects the boundary field through the coupling to the mass of scalar field. Then we restrain the field near the boundary $u = 0$, and therefore the term of $u^2$ inside the square brackets of Eq.(\ref{1-boundary-phi}) can be ignored safely. Meanwhile, we use a transformation $u = e^\theta$ and rewrite Eq.(\ref{1-boundary-phi}) in the following form
\begin{equation}\label{1-boundary-phi2}
 \frac{d^2 \phi}{d \theta^2} - 3 \frac{d \phi}{d \theta} - m^2 l^2 \phi = 0.
\end{equation}

We note that the motion equation (\ref{1-boundary-phi2}) is completely different from the usual AdS case \cite{Witten1,Gubser1,JMMaldacena}, since the timelike extra dimension is introduced here. There are some non-trivial information, not simply adding a dimension in space. So it is necessary to discuss the dimension of operators of boundary field corresponding to scalar field by using the above formulas. The characteristic equation is given by
\begin{equation}\label{1-character-eq}
\Delta \left(\Delta - 3\right) = m^2 l^2,
\end{equation}
where $\Delta$ are the dimensions of $\mathcal{O}_\phi$ duals to scalar field $\phi$.
The roots of Eq.(\ref{1-character-eq}) are given by
\begin{equation}\label{1-roots}
\Delta_{\pm} = \frac{3 \pm \sqrt{9 + 4 m^2 l^2}}{2}.
\end{equation}
It shows us an interesting phenomenon that the mass of field coupling with the extra dimension appeared in the dimensions of operator.

Then we check the Breitenlohner-Freedman boundary condition of $\phi$. In order to conveniently analyse the stabilization of AdS boundary vacuum, we define a stable extra dimension $l_{BF}$ satisfying Breitenlohner-Freedman boundary condition $m^2_{BF} l_{BF}^2 = -9/4$. We find when $m^2 l^2 > m^2_{BF} l_{BF}^2$, the vacuum of boundary AdS is stable. If the roots of the characteristic equation (\ref{1-character-eq}) are not equal with each other, we can have $\Delta_+ \neq \Delta_-$, thus
\begin{equation}\label{1-unequsolutions}
\phi = A(x) u^{\Delta_+} + B(x) u^{\Delta_-},
\end{equation}
where $\Delta_{\pm}$ could be real or complex numbers. For the complex numbers case, we have \begin{equation}\label{1-complexnum}
\Delta_{\pm} = \frac{3}{2} \pm i \sqrt{-\left(\frac{9}{4} + m^2 l^2\right)} \equiv \Delta_1 \pm i \Delta_2,
\end{equation}
where the behaviors of the scalar field is
\begin{equation}\label{1-com-field}
\phi = u^{\Delta_1} \left[A(x) u^{i \Delta_2} + B(x) u^{-i \Delta_2}\right],
\end{equation}
with the condition of $m^2 l^2 < -9/4$. Because it is oscillatory solution with enveloping line $u^{\Delta_1}$, the vacuum of boundary AdS with this kind solution is unstable. If $m^2 l^2 = m^2_{BF} l_{BF}^2$, it corresponds to the degenerate states with $\Delta_+ = \Delta_- = 3/2$. The solutions are given by
\begin{equation}\label{1-degen-states}
\phi(u, x) = \left[A(x) + B(x) \ln u\right] u^{\Delta}.
\end{equation}

After obtaining the Breitenlohner-Freedman boundary, we can give the field/operator correspondence for this scalar field. If $\Delta_+$ and $\Delta_-$ are both real and they are not equal with each other, the boundary field can be derived by the second term in Eq.(\ref{1-unequsolutions}), in which the coefficient $B(x)$ is the source of the operator $\mathcal{O}$. Thereby, a nonzero $B(x)$ leads directly to a interaction term
in the Lagrange of boundary field $\delta S_{boundary} = \int d^4 x B(x) \mathcal{O}(x)$. The coefficient $A(x)$ of the first term in Eq.(\ref{1-unequsolutions}) could be treated as the expected value of $\mathcal{O}$ shown by
\begin{equation}\label{1-oexpectedvalue}
< \mathcal{O}(x) > = \sqrt{9 + 4 m^2 l^2} A(x).
\end{equation}
The regular solutions for $B(x) = 0$ and $A(x) \neq 0$ mean that operator $\mathcal{O}$ spontaneously generates a expected value without source. In the momentum space, the correlation function is written as $G_R(\omega, k) = \sqrt{9 + 4 m^2 l^2} A(\omega, k)/B(\omega, k)$. If we invoke the incoming-wave boundary condition of Minkowski space correlators \cite{DTSon}, we can get the retarded Green's functions. Different boundary conditions could give us different quantization schemes. The condition of $B(x) = 0$ corresponds to the standard quantization and $A(x) = 0$ to an alternative quantization. The solution $u^{\Delta_+}$ is the normalizable mode and $u^{\Delta_-}$ is non-normalizable mode. According to the requirements of unitarity, the dimensions should satisfy the condition $\Delta > (d - 2)/2$ \cite{IRKlebano}. Hence, if $m^2 l^2 > -5/4$, $\Delta_-$ will not satisfy the above condition which will allow only one boundary condition of $B = 0$. However, for the case $-9/4 < m^2 l^2 < -5/4$, both condition $B = 0$ and $A = 0$ are allowed. If we add the Dirichlet or Neumann conditions to boundary, we can obtain two kind quantization schemes. For the alternative quantization, we can exchange $A$ and $B$ in Eq.(\ref{1-unequsolutions}). So $A(x)$ corresponds to the source and $B(x)$ to the expected value, and $\Delta_-$ is the conformal dimension of operator $\mathcal{O}$.

Then one may ask whether a finite temperature field theory exists in this AdS gravity? We will find after using a scaling about $t$ and $u$, i.e. $t \rightarrow t/L$ and $u \rightarrow L u$, a form of being asymptotic to AdS at the boundary could be obtained via a rotational transformation $l^2 = e^{2i z/u}$,
\begin{equation}\label{2-5dadssolution2}
d s_{\star}^2 = \frac{l^2}{u^2}\left[\mathcal{F}(u) d t^2 - \frac{d u^2}{\mathcal{F}(u)} - d x^2 - d y^2 - d z^2\right],
\end{equation}
where $\mathcal{F}(u) = 1 - u^3/u^3_0$ with horizon $u_0^3 = 1/ML^2$. Hence, when $u \rightarrow 0$, $\mathcal{F}(u) \rightarrow 1$, the boundary field indeed corresponds to the finite temperature and finite chemical potential case in which the metric satisfies the scaling invariance. Near the boundary we use a Wick rotation about $(u, x, y, z)$, metric (\ref{2-5dadssolution2}) reduces to a familiar Euclidean version of AdS,
\begin{equation}\label{2-euclideanads5}
d s^2 = \frac{l^2}{u^2} \left(d u^2 + d \vec{x}^2\right).
\end{equation}

In the end of this subsection, we will make a connection with already established results for
AdS and check whether our results match that of the usual case. In order to facilitate the comparison with the usual AdS case, we rewrite the metric (\ref{1-boundarymetric}) in the form of
\begin{equation}\label{add-euc-metric}
d s^2 = \frac{1}{u^2} \left[\frac{l^2}{L^4} d t^2 + l^2 d u^2 + \frac{l^2}{L^2} \left(d x^2 + d y^2\right)^2 + u^2 d l^2\right].
\end{equation}
Then comparing with the known Euclidean version of usual $AdS_{d+1}$ space shown by
\begin{equation}\label{add-usual-ads}
d s^2 = \frac{1}{u^2} \left[d \mathcal{T}^2 + d u^2 + \sum_{i =1}^{d - 1}\left(d x^i\right)^2\right],
\end{equation}
one can find extra dimension $l$ appears in all dimensions. For a given hypersurface of $l = constant$, metric (\ref{add-euc-metric}) will reduce to the usual case with $d = 3$ in (\ref{add-usual-ads}) when we adopt a certain coordinate transformations. According to the results of known $AdS$ case in Refs.\cite{BF1,BF2}, the scalar field $\Phi$ in the Euclidean version of the AdS background metric (\ref{add-usual-ads}), which can be looked as the analogy of $\phi$ in our calculation, has the following form
\begin{equation}\label{add-boundary-Phi}
u^2 \frac{\partial^2 \Phi}{\partial u^2} - \left(d - 1\right) u \frac{\partial \Phi}{\partial u} - \left[m^2 + (k^2 + \omega^2) u^2\right] \Phi = 0.
\end{equation}

Comparing Eq.(\ref{1-boundary-phi}) and Eq.(\ref{add-boundary-Phi}), we can find $\phi$ matches $\Phi$ with $d = 3$ on the hypersurface $l = constant$ when the extra dimension is absorbed into particles' mass and momentum through $m^2 l^2 \rightarrow m^2$, $L^2 k^2 \rightarrow k^2$ and $L^4 \omega^2 \rightarrow \omega^2$. Furthermore, near the boundary, Eqs.(\ref{1-boundary-phi}) and (\ref{add-boundary-Phi}) all induce to the same equation as Eq.(\ref{1-boundary-phi2}) with above scaling $m^2 l^2 \rightarrow m^2$. So we can say that if the extra dimension is constant in 5D, the boundary field near $u = 0$ in Eq.(\ref{1-boundary-phi}) will well match the usual 4D AdS case \cite{BF1,BF2}.
\subsection{two-points correlation functions}\label{sub-twopoint}
In this subsection, we consider a free scalar field with mass $m$ propagating in it, in which a massless scalar is dual to a marginal operator. It should be noticed that the operator here based on the duality of $AdS_4/CFT_3$. However, like the former subsection the AdS part in metric (\ref{1-boundarymetric}) contains the extra dimensional factor $l^2/L^2$, therefore we should consider the operator in the whole 4D field near the boundary of $AdS_4$ (see Fig.\ref{Fig1}). The extra dimension lives in the whole space, the dimensions of the field should be the (3 + 1) D field by adding one extra dimension.

Then we can build the boundary to bulk propagator $G(u, x; 0, x')$ which gives the bulk field configuration $\phi (u, x)$ for smooth boundary $\phi (0, x)$,
\begin{equation}\label{3-phi-1}
\phi (u, x) = \int d^4 x' G(u, x; 0, x') \phi (0, x),
\end{equation}
where $G$ is the retarded Green's function. In the Fourier space $k = (\omega, \textbf{k})$, the $\phi (u, x)$ has following form
\begin{equation}\label{3-phi-wan}
\tilde{\phi}(u, k) = \tilde{G} (u, k) \tilde{\phi}(0, k),
\end{equation}
where $\tilde{G} (u, k)$ is the Fourier space's solution to the mode equation of Eq.(\ref{1-motioneq}) as following
\begin{equation}\label{3-green-solution}
\frac{\partial^2 \tilde{G}}{\partial u^2} - \frac{2}{u} \frac{\partial \tilde{G}}{\partial u} - \left[\frac{m^2 l^2}{u^2} + k^2\right] \tilde{G} = 0,
\end{equation}
where after using the scaling $k \rightarrow k/L$ and $\omega \rightarrow \omega/L^2$, we can write $L^2 k^2 + L^4 \omega^2 \rightarrow k^2$ in momentum space. The boundary conditions are $\tilde{G} (u, k)|_{u\rightarrow 0} = 1$ and $\tilde{G} (u, k)|_{u\rightarrow \infty} = \text{finite}$. If we adopt a transformation $w(z)=u^{-3/2} \tilde{G}(u)$ with $z=k u$ and set a new parameter, i.e. $\nu^2 = 9/4 + l^2 m^2$, one can find the mode equation (\ref{3-green-solution}) reduces to a modified Bessel equation as followings,
\begin{equation}\label{3-virtual-Bessel-eq}
z^2 \frac{\partial^2 w}{\partial z^2} + z \frac{\partial w}{\partial z} - \left(z^2 + \nu^2\right) w = 0,
\end{equation}
where the solutions are the modified Bessel functions $I_{\pm \nu}(z)$ and $K_{\nu} (z)$, which are all regular functions throughout the $z$-plane cutting alone the negative real axis. So the solutions to Eq.(\ref{3-green-solution}) are $\tilde{G} = u^{3/2} I_{\pm \nu}(k u)$ and $\tilde{G} = u^{3/2} K_\nu (k u)$. Like the standard notation, the conformal weight of the operator $\mathcal{O}$ is shown by $\Delta = 3/2 + \nu$. The second solution is selected by the requirement of regularity in the interior. $I_{\pm \nu}(k u)$ increases exponentially as $u \rightarrow \infty$ and does not lead to a finite action configuration. Hence, the solution regular at $u = \infty$ and equal one at $u = \epsilon$ is given by
\begin{equation}\label{3-greensolution}
\tilde{G} (u, k) = \frac{u^{3/2} K_\nu (k u)}{\epsilon^{3/2} K_\nu (k \epsilon)}.
\end{equation}

Then by standard integration by parts, we can obtain the on-shell bulk action determined by the boundary field,
\begin{eqnarray}\label{3-boundaryfield}
\nonumber S &=& \int \frac{d^4 k d^4 k'}{(2\pi)^8} (2\pi)^4 \delta^4 (k + k')\phi_0(k) \phi_0(k')\frac{l^2 \tilde{G} (k')\partial_u \tilde{G} (k)}{u^2 L^4}\bigg|_{\epsilon}^{\infty}\\ &=& \int \frac{d^4 k d^4 k'}{(2\pi)^8} \phi_0(k) \phi_0(k') \mathcal{F} (u, k ,k') \bigg|_{\epsilon}^{\infty},
\end{eqnarray}
where the flux factor $\mathcal{F}$ is given by
\begin{equation}\label{3-flux-factor}
\mathcal{F} (u, k ,k') = \tilde{G} (u, k') \sqrt{g} g^{uu} \partial_u \tilde{G} (u,k).
\end{equation}
It only receives the contribution from the cutoff at $u = \epsilon$ because the propagator vanishes at large $u$. Hence differentiating $S[\phi]$ with respect to $\phi_0 (k')$ and $\phi_0 (k)$, one can obtain two-points function for the operator $\mathcal{O}$ dual to $\phi$,
\begin{equation}\label{3twopointfunction}
<\mathcal{O}(k')\mathcal{O}(k)> = Z^{-1} \frac{\delta^2 Z[\phi_0]}{\delta \phi_0 (k) \delta \phi_0 (k')} \bigg|_{\phi_0 = 0}= - 2 \mathcal{F} (z, k, k')\bigg|_{\epsilon},
\end{equation}
where $Z [\phi]$ is the partition function defined in Ref. \cite{Witten1,DTSon}. Then if we substitute Eq.(\ref{3-boundaryfield}) into the flux factor $\mathcal{F}$, one can obtain the two-points function in a form of series as
\begin{equation}\label{3twopointfunctionseries}
<\hat{\mathcal{O}}(k')\hat{\mathcal{O}}(k)> = - \frac{l^2(2\pi)^4}{L^4} 2^{4 - 2\Delta} \epsilon^{2(\Delta - 3)} \delta^4 (k + k') k^{2\Delta - 3} \frac{\Gamma (\frac{5}{2} - \Delta)}{\Gamma (\Delta - \frac{3}{2})} + \cdots,
\end{equation}
where the leading analytic terms in $(\epsilon k)^2$ and the higher order terms in $(\epsilon k)^2$ are not listed. For the massless case, i.e. $\Delta = 3$, we can have
\begin{equation}\label{3masslesstwopoint}
<\hat{\mathcal{O}}(k')\hat{\mathcal{O}}(k)> = \frac{l^2(2\pi)^4}{L^4} \delta^4 (k + k') k^3.
\end{equation}

At the last, we will compare our results with the usual Minkowski space in Euclidean version \cite{DTSon}. We can find three main points are different as followings. Firstly, the mode equation is different. The mode equation of usual Minkowski case given in Ref.\cite{DTSon} is written as the following form
\begin{equation}\label{add-model-eq}
\frac{\partial^2 \tilde{G}}{\partial u^2} - \frac{3}{u} \frac{\partial \tilde{G}}{\partial u} - \left[\frac{m^2 R^2}{u^2} + k^2\right] \tilde{G} = 0,
\end{equation}
where $R$ is the curvature. Comparing Eqs.(\ref{add-model-eq}) and (\ref{3-green-solution}), we can find just the first order differential term and particle mass term are different. Because the scalar field equation (\ref{1-motioneq}) is determined by the space (\ref{1-boundarymetric}). The 5D used here is Ricci flat and the curvature of 4D induced AdS metric is determined by extra dimension $l$. The mode equation of field (\ref{3-green-solution}) drives us to get the results about two-points correlation functions that are different from usual Minkowski case.

Secondly, the orders $\nu$ in the boundary field is different. Although the mode equations (\ref{add-model-eq}) and (\ref{3-green-solution}) of boundary field can be reduced into the same type of a modified Bessel equation, the orders $\nu$ have different magnitudes, i.e. $\nu^2 = 9/4 + l^2 m^2$ for this Ricci flat case and $\nu^2 = 4 + m^2 R^2$ for Minkowski case.

Thirdly, the conformal weight $\Delta$ of the operator $\mathcal{O}$ is different, namely $\Delta = \nu + 3/2$ for our case and $\Delta = \nu + 2$ for the Minkowski case. It clearly shows that the timelike extra dimension can effect the two-points correlation functions both in massless and massive fields when comparing with the normal Minkowski space \cite{DTSon}.
The reason of these differences is that the 5D space is Ricci flat and the hypersurface $M^{4*}$ is the induced 4D AdS space. Because the 4D part is dependent on the extra dimension, the correspondence between $AdS_4$ and $CFT_3$ must be considered in the full 5D picture, which is illustrated in Fig.\ref{Fig1}.
\section{conclusion}\label{conclusion}
In this paper, a general AdS solution of 5D relativity is presented by solving the Kaluza-Klein equation $R_{AB} = 0$. The timelike extra dimension gives us a 4D negative cosmological constant and a 4D Lorentzian negative constant curvature space. These results are different from the spacelike case. The later offers us the dS solution with a 4D positive cosmology constant \cite{stm-reviews1,stm-reviews2,stm-reviews3,stm-reviews4}. We summarize what have achieved and make some further comments.

With the development of gravity, the AdS space has drawn more and more attention. It is believed that through a certain duality the AdS is connected with some quantum fields. The latter could be used to explain many phenomenological physics in the strongly coupled system. However, there is no work related to AdS solution in the 5D relativity \cite{stm-reviews4}. With the minimal hypothesis, we find the AdS solution and the negative cosmological constant are naturally obtained if the extra dimension is timelike. Meanwhile, we calculate the motion equation of a test particle and find the 4D motion could be geodesics if $\overset{\star}{g}_{\alpha\beta} = 0$. The solution of the part of extra dimension (\ref{fifthgeo-so}) is trigonometric function, which is unlike the Hyperbolic functions in positive cosmological constant \cite{cc-BMashhoon1}. Our calculations also indicate that the 5D relativity give the same form of the extra force $F^{\mu}$ no matter in dS case \cite{stm-reviews1,stm-reviews2,stm-reviews3} or in AdS case. The parallel part $p^{\mu}$ in $F^{\mu}$ will not disappear unless the metric satisfies the condition of $\overset{\star}{g}_{\alpha\beta} = 0$.

In the AdS solution obtained here, one can find if the extra dimension is timelike, as opposed to be spacelike, the holographic principle can be used on the brane. Both choices of whether extra dimension is timelike or spacelike are allowed because the timelike extra dimension does not have the physical nature of a time \cite{wavelikes1}. After a simple implication of holographic duality, we obtain the field/operator corresponding. Comparing with the case of higher dimensional nonzero cosmological constant shown by Witten \cite{Witten1}, we will find the mass of field coupling with the extra dimension appeared in the dimensions of operator. If $m^2 l^2 > -5/4$, the boundary condition of $B = 0$ is accepted for the unitarity of scalar field. But if $-9/4 < m^2 l^2 < -5/4$, the conditions $B = 0$ and $A = 0$ are allowed. So after adding the Dirichlet or Neumann conditions to boundary, one can get two quantization schemes. Meanwhile, we also compute the retarded Green's functions by the prescription of flux factor $\mathcal{F}$: $G^R (k) = -2 \mathcal{F} (k, z)\big|_{zB}$ \cite{DTSon}. The two-points correlation functions are shown in Eq.(\ref{3twopointfunctionseries}) for massive scalar, and in Eq.(\ref{3masslesstwopoint}) for massless case. It indicates the timelike extra dimension can effect the transport property of linear response theory in the region on the boundary.

Finally, we need to clarify the holographic duality used here, and two points need us to pay attention. (i)
Because the 5D relativity is Ricci flat, there is no 5D $AdS_5$ solution to the Kaluza-Klein equations $R_{AB} = 0$. The usual 4D negative cosmological constant $\Lambda$ is induced from the extra dimension, and the $AdS_4$ thus lives on the hypersurface $M^{4*}$. (ii) There is a coupling factor, i.e. $l^2/L^2$ which comes from the extra dimension, before the part of 4D AdS in Eq.(\ref{1-boundarymetric}). Therefore, the method of directly cutting the AdS part from higher dimensional space can not available for the 5D Ricci flat canonical system. Based on the points (i) and (ii), in order to implicate the the holographic duality we adopt a conservative scheme in the way of that the internal key duality is still $AdS_4/CFT_3$ when considering the point (i), but the holographic duality should be calculated in the whole 5D space when considering the point (ii). So, we should add the extra dimension $l$ into the field near the boundary of $AdS_4$. The dimension of the field consists of the 3 D from $AdS_4/CFT_3$ and the 1 D from the extra dimension. In the calculations, we all start from the 5D space (\ref{1-boundarymetric}), and obtain the field/operator corresponding near the boundary of $AdS_4$ in the subsection \ref{sub-fieldoperator} and the two-points correlation functions in subsection \ref{sub-twopoint}. The boundary to bulk propagator $G(u, x; 0, x')$ give us the nonsingular bulk field configuration $\phi (u, x)$ for any smooth boundary value $\phi (0, x)$ near the boundary of $AdS_4$ (see Fig.(\ref{Fig1})). The mode equation (\ref{3-virtual-Bessel-eq}) of Green's function $G$ is derived from the boundary field theory in subsection \ref{sub-fieldoperator}.

\acknowledgments
We thank the anonymous referee for helpful corrections. We also need express our sincere and profound thanks to Prof. P. S. Wesson for his encouragement on the extra dimension gravitational theory and the useful discussion about the initial motivation of this paper. We also need thank Dr. Hongbao Zhang for useful discussions. This work is supported by the National Natural Science Foundation of China under grants 11475143, Science and Technology Innovation Talents in Universities of Henan Province under grant 14HASTIT043, the Nanhu Scholars Program for Young Scholars of Xinyang Normal University.


\begin{thebibliography}{*}
\bibitem{Kaluza}T. Kaluza, Sitz. Preuss. Akad. Wiss. 33 (1921) 966.
\bibitem{Klein}O. Klein, Z. Phys. 37 (1926) 895.
\bibitem{stm-reviews4}P.S. Wesson, The status of modern five-dimensional gravity, Int. J. Mod. Phys. D. 24 (2015) 1530001, arXiv:1412.6136 [gr-qc].
\bibitem{stm-reviews1}J.M. Overduin and P.S. Wesson, Kaluza-Klein gravity, Phys. Rep. 283 (1997) 303, arXiv:gr-qc/9805018.
\bibitem{stm-reviews2}P.S. Wesson, Space-time-matter: modern Kaluza-Klein theory, World Scientific, Singapore, 1999.
\bibitem{stm-reviews3}P.S. Wesson, Five-dimensional relativity, World Scientific, Singapore, 2006.
\bibitem{stm-apj1}D. Kalligas, P.S. Wesson and C.W.F. Everitt, The classical test in Kaluza-Klein, Astrophys. J. 439 (1995) 548.
\bibitem{stm-apj2}H. Liu and J. Overduin, Solar system tests of higher dimensional gravity, Astrophys. J. 538 (2000) 386, [arXiv:gr-qc/0003034].
\bibitem{stm-bigbounce1}H. Liu and B. Mashhoon, A machian interpretation of the cosmological constant, Ann. Phys. (Leipzig) 4 (1995) 565.
\bibitem{stm-bigbounce2}H. Liu, and P.S. Wesson, Universe models with a variable cosmological ``constant" and a ``big bounce", Astrophys. J. 562 (2001) 1, [arXiv:gr-qc/0107093].
\bibitem{QNM-BH}M. Liu, H. Liu and Y. Gui, Quasi-normal modes of massless scalar field around the 5D Ricci-flat black string, Class. Quant. Grav. 25 (2008) 105001, arXiv:0806.2716 [gr-qc].
\bibitem{entropy1-BH}M. Liu and H. Liu, The dynamic behavior of quantum statistical entropy in 5D Ricci-flat black string with thin-layer approach, Phys. Lett. B 661 (2008) 365-369, arXiv:0811.0655 [gr-qc].
\bibitem{entropy2-BH}M. Liu, Y. Gui and H. Liu, Quantum statistical entropy and minimal length of 5D Ricci-flat black string with generalized uncertainty principle, Phys. Rev. D 78 (2008) 124003, arXiv:0812.0864 [gr-qc].
%braneworld
\bibitem{ADD1}N. Arkani-Hamed, S. Dimopoulos and G. Dvali, The hierarchy problem and new dimensions at a millimeter, Phys. Lett. B 429 (1998) 263, [arXiv:hep-ph/9803315].
\bibitem{ADD2}I. Antoniadis, N. Arkani-Hamed, S. Dimopoulos and G. Dvali, New dimensions at a millimeter to a Fermi and Superstrings at a TeV, Phys. Lett. B 436 (1998) 257, [arXiv:hep-ph/9804398].
\bibitem{RS1}L. Randall and R. Sundrum, A large mass hierarchy from a small extra dimension, Phys. Rev. Lett. 83 (1999) 3370, [arXiv:hep-ph/9905221].
\bibitem{RS2}L. Randall and R. Sundrum, An alternative to compactification, Phys. Rev. Lett. 83 (1999) 4690, [arXiv:hep-th/9906064].
\bibitem{RMaartens}R. Maartens and K. Koyama, Brane-world gravity, Living Rev. Relativity 13 (2010) 5, [arXiv:1004.3962[hep-th]].
\bibitem{GDvali}G. Dvali, G. Gabadadze and G. Senjanovic, Constraints on extra time dimensions, [arXiv:hep-ph/9910207].
%space-like dimension model
\bibitem{YShtanov}Y. Shtanov and V. Sahni, Bouncing braneworlds, Phys. Lett. B557 (2003) 1, [arXiv:gr-qc/0208047].
\bibitem{MChaichian}M. Chaichian and A.B. Kobakhidze, Mass hierarchy and localization of gravity in extra time, Phys. Lett. B 488 (2000) 117, [arXiv:hep-th/0003269].
\bibitem{ZBerezhiani}Z. Berezhiani, M. Chaichian, A.B. Kobakhidze and Z.H. Yu, Vanishing of cosmological constant and fully localized gravity in a brane world with extra time(s), Phys. Lett. B 517 (2001) 387, [arXiv:hep-th/0102207].
\bibitem{YVShtanov}Y.V. Shtanov, Closed system of equations on a brane, Phys. Lett. B 541 (2002) 177, [arXiv:hep-ph/0108153].
%add
\bibitem{tHooft}G. 't Hooft, Dimensional reduction in quantum gravity, [arXiv:gr-qc/9310026].
\bibitem{LSusskind}L. Susskind, The world as a hologram, J. Math. Phys. 36 (1995) 6377, [arXiv:hep-th/9409089].
\bibitem{JMMaldacena}J.M. Maldacena, The large N limit of superconformal field theories and supergravity, Adv. Theor. Math. Phys. 2 (1998) 231, [arXiv:hep-th/9711200].
\bibitem{OAharony}O. Aharony, S.S. Gubser, J. Maldacena, H. Ooguri and Y. Oz, Large N field theories, string theory and gravity, Phys. Rept. 323 (2000) 183, [arXiv:hep-th/9905111].
%others
\bibitem{Witten1}E. Witten, Anti de Sitter space and holography, Adv. Theor. Math. Phys. 2 (1998) 253, [arXiv:hep-th/9802150].
\bibitem{Gubser1}S.S. Gubser, I.R. Klebanov and A.M. Polyakov, Gauge theory correlators from non-critical string theory, Phys. Lett. B 428 (1998) 105, [arXiv:hep-th/9802109].
\bibitem{IRKlebano}I.R. Klebano and E. Witten, AdS/CFT correpondence and symmetry breaking, Nucl. Phys. B 556 (1999) 89.
\bibitem{DTSon}D.T. Son and A.O. Starinets, Minkowski-space correlators in AdS/CFT correspondence: recipe and applications, JHEP 0209 (2002) 042, [arXiv:hep-th/0205051].
%wavelike
\bibitem{wavelikes1}P.S. Wesson, Vacuum waves, Phys. Lett. B 722 (2013) 1, [arXiv:1301.0333[physics.gen-ph]].
\bibitem{wavelikes2}P.S. Wesson and J.M. Overduin, Waves and causality in higher dimensions, Phys. Lett. B 750 (2015) 302, [arXiv:1510.07178 [gr-qc]].
\bibitem{wavelikes3}P.S. Wesson, The dispersion relation for matter waves in a two-Phase vacuum, Mod. Phys. Lett. A 29 (2014) 1450168, [arXiv:1411.0046 [gr-qc]].
%canonical coordinates
\bibitem{cc-BMashhoon1}B. Mashhoon, H.Y. Liu and P.S. Wesson, Particle masses and the cosmological constant in Kaluza-Klein theory, Phys. Lett. B 331 (1994) 305.
\bibitem{cc-PSWesson1}P.S. Wesson, B. Mashhoon, H.Y. Liu and W.N. Sajko, Fifth force from fifth dimension, Phys. Lett. B 456 (1999) 34.
\bibitem{cc-HYLiu}H.Y. Liu and B. Mashhoon, Spacetime measurements in Kaluza-Klein gravity, Phys. Lett. A 272 (2000) 26.
\bibitem{cc-BMashhoon2}B. Mashhoon and P.S. Wesson, Gauge-dependent cosmological ``constant", Class. Quant. Grav. 21 (2004) 3611, [arXiv:gr-qc/0401002].
\bibitem{LDLandau}L.D. Landau and E.M. Lifshitz, The classical theory of fields, Pergamon, Oxford, 1975.
\bibitem{Horowitz_PRD}G. T. Horowitz and M. M. Roberts, Holographic Superconductors with Various Condensates, Phys. Rev. D 78 (2008) 126008, arXiv:0810.1077 [hep-th].
\bibitem{Hartnoll_PRL}S. A. Hartnoll, C. P. Herzog and G. T. Horowitz, Building an AdS/CFT superconductor, Phys. Rev. Lett. 101 (2008) 031601, arXiv:0803.3295 [hep-th].
\bibitem{Hawking}S. Hawking and G.F.R. Ellis, The large scale structure of space-time, Cambridge University Press, 1973.
\bibitem{liangzhou}C. B. Liang and B. Zhou, Introduction of Differential Geometry and General Relativity, China Science Publishing, Beijing, 2009.
\bibitem{Abramowitz}M. Abramowitz and I. Stegun, Handbook of mathematical functions, Dover, New York, 1972.
\bibitem{BF1}P. Breitenlohner and D.Z. Freedman, Positive energy in Anti-de Sitter backgrounds and gauged extended supergravity, Phys. Lett. B 115 (1982) 197.
\bibitem{BF2}P. Breitenlohner and D.Z. Freedman, Stability in gauged extended supergravity, Annals Phys. 144 (1982) 249.
\end{thebibliography}
\end{document}